\documentclass[submission,copyright,creativecommons]{eptcs}
\providecommand{\event}{WCSI10} 
\usepackage{makeidx}  
\usepackage[T1]{fontenc}   
\usepackage{amsmath,amssymb}
\usepackage{xcolor}
\usepackage{tikz}
\usepackage{ae}
\usepackage{xspace}
\usepackage{boxedminipage}
\usepackage{url}
\usepackage{multicol}
\usepackage[savemem]{listings}
\usepackage{wrapfig}

\definecolor{colKeys}{rgb}{0,0,1}
\definecolor{colIdentifier}{rgb}{0,0,0}
\definecolor{colComments}{rgb}{0,0.5,1}
\definecolor{colString}{rgb}{0.6,0.1,0.1}
\definecolor{violet}{rgb}{0.5,0,0.5}

\def\repSPEC{./SPECS}
\def\repFIGURES{./FIGURES}

\def\kmelia{\textsf{Kmelia}\xspace}
\def\Lotos{\textsc{Lotos}\xspace}

\newcommand{\ie}{\textit{i.e.}}

\def\beta{\mathbb{B}}

\def\kmelia{\textsf{Kmelia}\xspace}
\def\key{\textsf{Key}\xspace}
\def\B{\textsf{B}\xspace}

\newcommand{\repSTYLES}{./STYLES}
\usepackage{\repSTYLES/Kmelia}
\newcommand{\K}[1]{\lstinline[language=Kmelia,basicstyle={\small \sffamily}]{#1}}
\usepackage{\repSTYLES/B2}
\newcommand{\BN}[1]{\lstinline[language=Bmethod,basicstyle={\rmfamily \small}]{#1}}
\newcommand{\JJ}[1]{\lstinline[language=java,basicstyle={\rmfamily \small}]{#1}}
\newcommand{\specKmelia}[4] {\lstinputlisting[language=Kmelia,
  basicstyle={\tiny \sffamily}, frame=trBL, framesep=2pt,
  caption={\kmelia specification \K{#1}}, label={#2}, linerange={#4}]{#3}}
\newcommand{\specKmeliaFirst}[4] {\lstinputlisting[language=Kmelia,
  basicstyle={\tiny \sffamily}, frame=trL, framesep=2pt,
  caption={\kmelia specification \K{#1}}, label={#2}, linerange={#4}]{#3}}
\newcommand{\specKmeliaSecond}[2] {\lstinputlisting[language=Kmelia,
  basicstyle={\tiny \sffamily}, frame=rBL, framesep=2pt,
  linerange={#2}]{#1}}
\def\repFIGURES{./FIGURES}
\def\titlerunning{Multilevel Contracts for Trusted Components}
\def\authorrunning{M. Messabihi, P. André  \& C. Attiogbé}

\title{Multilevel Contracts for Trusted Components}
\author{\and
Mohamed Messabihi \qquad Pascal Andr\'e  \qquad Christian Attiogb\'e
\institute{LINA UMR CNRS 6241 \\ ~~University of Nantes, France}
\email{FirstName.LastName@univ-nantes.fr}
}

\begin{document}
\maketitle
\begin{abstract}
This article contributes to the design and the verification of trusted components and services.
The contracts are declined at several levels to cover then different facets, such as component consistency, compatibility or correctness.
The article introduces  multilevel contracts and a design+verification process
for handling and analysing these contracts in component models.
The approach is implemented with the COSTO platform that supports the \kmelia component model.
A case study illustrates the overall approach.
\end{abstract}

\section{Introduction}
\label{intro}
Component-Based Software Engineering (CBSE) using \emph{off-the-shelf} components is one approach to deal with the software complexity.
Since the components may be developed by third-parties, assembling them
requires  means to ensure the correctness of the component behaviours and their interoperability.
This requires first that the components have rich interface descriptions and second the availability of a verification process to check the given properties.
In our work we tackle the issue of building trusted components by the means of \emph{contracts}.

As a component is usually defined as \emph{"an unit of composition with \textbf{contractually} specified interfaces and explicit context dependencies only"} \cite{Szyperski2002}, the notion of contract appears to be a natural solution to express and to organise component specification and verification. 
The contracts are suitable to express properties such as the component's consistency preservation or the component interoperability.
By contract we mainly refer to contract-based design \cite{Meyer92applyingdesign} which extended the use of Hoare assertions (pre/post-conditions, invariant) at design and programming levels. 
But contracts may have different meanings depending on the context (component or assembly of components) or the facet (signature, structure, assertions, dynamics) \cite{DBLP:conf/cbsq/ReussnerPS03,Beugnard,brogiJLAP2010}.
Therefore we use the term \emph{multilevel contract} in this work.

In order to improve the confidence of the components and their assemblies, it is necessary to make contracts explicit~\cite{Beugnard}. 
This demands a strong emphasis on their analyzability
early in development process and a way to ensure systematically the correctness of the components with respect to the contracts. However, most of today component-based technologies lack formal analysis tools to ensure the component dependability.
Our work contributes in satisfying this need.

The contribution of this article is as follows.
We consider a multilevel contract approach which covers a specification activity and an analysis activity realised by verification techniques.
We show (i) how the contracts can be defined at different levels (service, component, assembly) of a component model in order to specify several kinds of correctness properties
and (ii) how the contracts can be checked.
The demonstration applies to our experimental \kmelia component model and its language.
The core of the data language of \kmelia is a first order logic, it is extended with user-defined data types and related statements; the behaviour language is based on transition systems.   
We experiment this proposal with the \textsf{COSTO} (COmponent Study TOolkit), a toolbox associated to the \kmelia model. Yet \textsf{Kmelia/COSTO} enables us to experiment at an abstract level, with medium size systems with intensive data and client-provider interaction style.

The remainder of the article is organised as follows.
In Section~\ref{correctness} we introduce the multilevel contract approach; the considered levels and the contracts are made explicit. 
In Section~\ref{section:process} we overview the global design and verification process based on the multilevel contracts.
Our contract approach is implemented in the \kmelia component model; Section~\ref{kmelia} describes how contracts are integrated at different levels. 
The support case study is a simple bank Automatic Teller Machine \textsf{(ATM)}.
Section~\ref{verification} illustrates the verification process in the case of  \kmelia and the experimentations on the ATM case study.
In Section \ref{related} we discuss related approaches.
Finally Section~\ref{conclusion} concludes the article and describes planed extensions of this work.

\section{Using Multi-levels Contracts in Component Models}
\label{correctness}
%

In this section we assume a general service component model where a component interface is defined by one or several services expressing provided or required functionalities; a component may be assembled with other components via its interface; a component may have invariant properties; a service may have a dynamic behaviour including interactions with other service components.
According to~\cite{meyer03grand}, \textit{a Trusted  Component  is  a  reusable  software element possessing specified and guaranteed property qualities}. 
The notion of contract is helpful to model various kind of properties.
Contracts take place in  services as assertions (pre/post-conditions), 
in components as invariants to preserve by the services, or 
in assemblies as component compatibility properties.
These general contracts should be made precise and extended to cope with the expressiveness of the considered component models.
In particular the \textit{interoperability} between components should consider the following properties. 
\begin{itemize}
\item
\textit{Static interoperability properties}:
the compatibility of interface signatures (naming and typing);
does a component give enough information about its interface(s) in order to be (re)usable by other components ?
\item
\textit{Architectural properties}: 
the availability of the required components,
the availability of the required services,
the correctness of the linked component interfaces;
\item
\textit{Functional properties}:
do the components do what they must do?
These correctness properties may be checked both on each component and on the component assemblies and compositions.
\item
\textit{Behavioural compatibility}:  the correct interaction between two or more components which are combined.
The properties depends on the interaction model features: sequential vs. concurrent, call vs. synchronisations, synchronous vs asynchronous, pair vs. multipart communication, shared data, atomic/structured actions...
\end{itemize}

In order to cope with different meaning and different context, we introduce the notion of multilevel contract.
A \textit{multilevel contract} is a contract defined at different structure level 
(service, component, assembly, composition)
according to different expected properties.
The hierarchical vision of the contracts provides a convenient framework to master the incremental building of components and the latter verification process.
In the following, we detail the main properties associated to each level.

\paragraph{Service contract}
A contract at the service level expresses that the service terminates in a consistent state. This contract deals with the \textit{behavioural consistency} and  the \textit{functional correctness} properties.
\begin{itemize}
\item The \textit{behavioural consistency} property states that the execution of the service actions  does not lead to inconsistent states (such as deadlock).
\item The \textit{functional correctness} property expresses that
a service achieves what it is supposed to do.
The functional correctness of a service of a component is defined here using the Hoare-style specification (Pre-condition, Statement, Post-condition) where Statement is the service behaviour.
This property should be checked with respect to the requirements of the owner component.
\end{itemize}

\paragraph{Component contract}
At the component level the contract states that this component can be reused with confidence.
It deals with three properties: the \textit{component consistency}, the \textit{protocol correctness} property and also the \textit{service accessibility}. 
A component protocol is defined here as the set of all the valid sequences of service invocations.
\begin{itemize}
\item The  \textit{component consistency} property states that the invariant properties of the component are preserved by all the services embodied in the component.
Considering that a component equiped with services is \textit{consistent} if its properties are always satisfied  whatever the behaviour of the services is, one can set a consistency preservation contract between the services and their owner component to ensure that property.
\item The \textit{protocol correctness} property expresses that
the order in which the services are to be invoked by clients is correct with respect to the rules given by the services' specification.
\item The \textit{service accessibility} property states that the services defined in the interface of a component are available. This is related to intra-component traceability of service dependency.
\end{itemize}

\paragraph{Assembly contract}
In an assembly, made of linked trusted components, each component will contribute to the well-formedness of the links by requiring or ensuring appropriate assertions: this is the coarse-grained contract.
The link establishes a client/supplier relation.
The assembly contract covers correctness properties with four layers:
\begin{itemize}
\item The first layer deals with the \textit{service signature compatibility} among the services of the interfaces of the assembled components. 
The service call should respect the service signature.
The signature matching between the involved services of component interfaces covers at least name resolution, visibility rules, typing and subtyping rules. 

\item The second layer deals with the \textit{service structure consistency} of the assembled components. Assuming that services can be composed from other (sub)services, connecting services is possible only if their structures are compatible (but not necessary identical).

\item The third layer deals with the \textit{service compliance} of the assembled components.
If the services use a Hoare-like specification, one has to relate their pre-conditions and post-conditions \cite{zaremski97a}. The caller pre-condition is stronger than the called one. The called post-condition is stronger than the caller's one. Each part involved in the assembly should fulfil its counterpart of the contract.
\item The fourth layer deals with the \textit{behavioural compatibility} between the linked services of the assembled components.
Behavioural compatibility is about the correct interaction between two or more components which are combined through their services.
\end{itemize}
The following table summarises the crossing of levels and properties (layers) covered by the multilevel contracts.
Note that the composite level is not dealt with in this paper.

\footnotesize
\begin{center}
\begin{tabular}{|c|c|c|c|}
\hline
\textbf{service} & \textbf{component} & \textbf{assembly} & \textbf{composite}\\
\hline
\hline
behavioural consistency & component consistency & service signature compatibility (ssic) & ssic\\
\hline
functional correctness & protocol correctness & service structure consistency (sstc) & sstc\\
\hline
 & service accessibility & service compliance (sco) & sco \\
\hline
 & & behavioural compatibility (bhc) & bhc\\
\hline
\end{tabular} 
\end{center}
\normalsize

The four layers above are useful to define interoperability levels.
A Corba component with IDL interfaces can be compatible only at the first level with other models.
The four layers can be augmented with other kind of properties like the quality of service.

In  Section~\ref{section:process} we provide the details of the global verification process based on the above levels and contracts.
It applies to component models with high level services such as the \kmelia multiservices component model, which serves as a working context for applying the multilevel contract definition (Section~\ref{kmelia}) and its verification (Section~\ref{verification}).

\section{Multi-levels Contract Design and Verification Process}
\label{section:process}
%
%
This section presents a component-based design process which takes into account the multilevel contracts. 
The components and assemblies are assumed to be abstract, meaning that they are independent from execution platforms. They can be refined or implemented later in centralised or distributed execution platforms.

\begin{figure}[!ht]
\begin{center}
\includegraphics[width=0.8\linewidth]{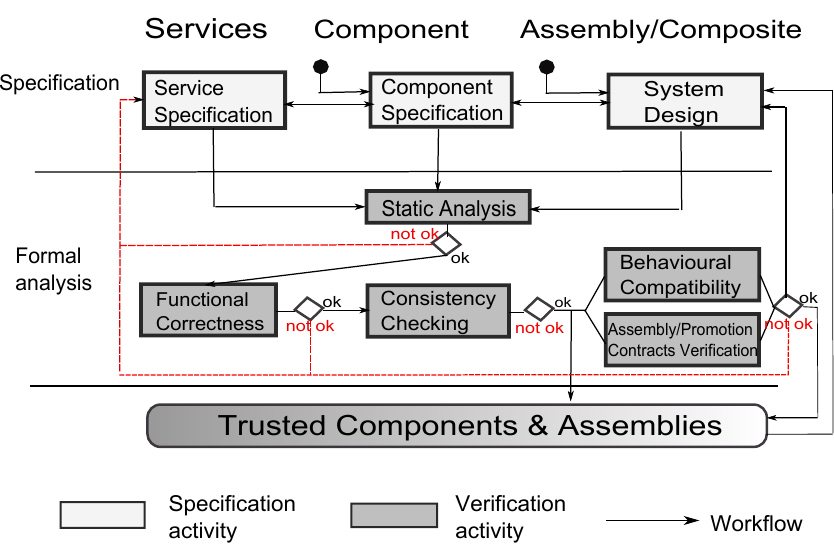}
 \caption{Contract design process}\label{process}
\end{center}
\end{figure}

As depicted in Figure~\ref{process}, the process is divided in two phases: 
the specification phase made of specification activities
and the formal analysis phase made of verification activities.
The workflow is presented as a whole but the activities can be performed iteratively in any order.
From a practical point of view, 
the specifier would switch from one phase to the other according to a customised methodology, inspired from top-down or bottom-up approaches, with a component or system orientation.
For example the specifier may iterate on the component level only to deliver components off the shelf.
This design approach allows the reuse of designed components by making the component descriptions available in a component library.

\subsection{Specification phase: making contracts explicit}

The specification phase includes three activities:
a software system design (assembly/composition),
a software component specification and a service specification.
In a top-down approach, the \emph{system design} activity starts first.
It defines the system as a collection of interacting subsystems and components. 
If components or assemblies that match the requirements already exist on the shelf, they can be directly integrated in the system design. Otherwise, the \emph{component specification} activity will produce the new component(s). Once the component structure is established, the detailed \emph{service specification} activity proceeds.
The main point in this phase is that the contracts must be explicitly expressed at each level in order to be checked.

\subsection{Formal analysis based on contract checking}

The models produced during the specification phase are analysed by checking the contracts.
The verification process iterates on five formal analysis activities as depicted in Figure~\ref{process}, each activity  refers to contracts of  Section~\ref{correctness}.
\begin{enumerate}
\item The \emph{Static analysis} activity checks the syntactic correctness at all levels, the service accessibility of the component level, and the \textit{static interoperability} of the assembly level, which itself covers the service signature compatibility and the service structure consistency.
\item The \emph{Functional correctness} activity checks the \textit{behavioural consistency} property at the service level and a part of the \textit{protocol correctness} property at the component level.
\item The \emph{Consistency checking} activity covers the \textit{component consistency} property at the component level.
\item The \emph{Behavioural compatibility} activity checks the \textit{behavioural consistency} property at the service level, a part of the \textit{protocol correctness} property at the component level and the \textit{behavioural compatibility} at the assembly level. 
\item The \emph{Assembly/Promotion contracts verification} activity checks the \textit{service compliance} of the assembled components at the assembly level.
The promotion consists in making a component feature available at the composite level \ie\ a component service can be promoted as a composite service possibly with restrictions or extensions.
The promotion is treated in the context of composite components, which is out of the scope of this article.
\end{enumerate}

We are not going to deal with all the details of the verification activities.
Section~\ref{verification} provides a concrete material on how the process is put into practice in the context of \textsf{COSTO/Kmelia}.
\section{Designing Contracts in the Kmelia Component Model}
\label{kmelia}
%
We introduce here the main features of \kmelia, an abstract and formal component model~\cite{sc06}; an up-to-date formal description of the model can be found in~\cite{facs09}. We illustrate the use of contracts with a simple bank Automatic Teller Machine (ATM).

The key features of \kmelia are:
\begin{itemize}
\item \textit{service}: a service describes a functionality; it is more than a simple operation; it has a pre-condition, a post-condition and a behaviour described with a labelled transition system (LTS). Moreover a service may hierarchically give access to other services. The behaviour supports communication interactions, dynamic evolution rules and service composition;
\item \textit{component}: a component is a container of services; it is described with a state space constrained by an invariant.
A component is  designed independently from its environment
by setting assumptions such as virtual client components or required service specifications;
\item \textit{assembly of components}: an assembly is a set of components linked via their required and provided services with the aim to build effective functionality. Linking components by their services in assemblies establishes a possible bridge to Service Oriented Architectures. The component assemblies are governed by strict service composability rules;
\item \textit{composite component:} a composite component is a component that encapsulates assemblies or other components; it is subject to encapsulation and promotion policies.
\end{itemize}

\begin{figure}[!ht]
\centerline{\includegraphics[width=0.85\textwidth]{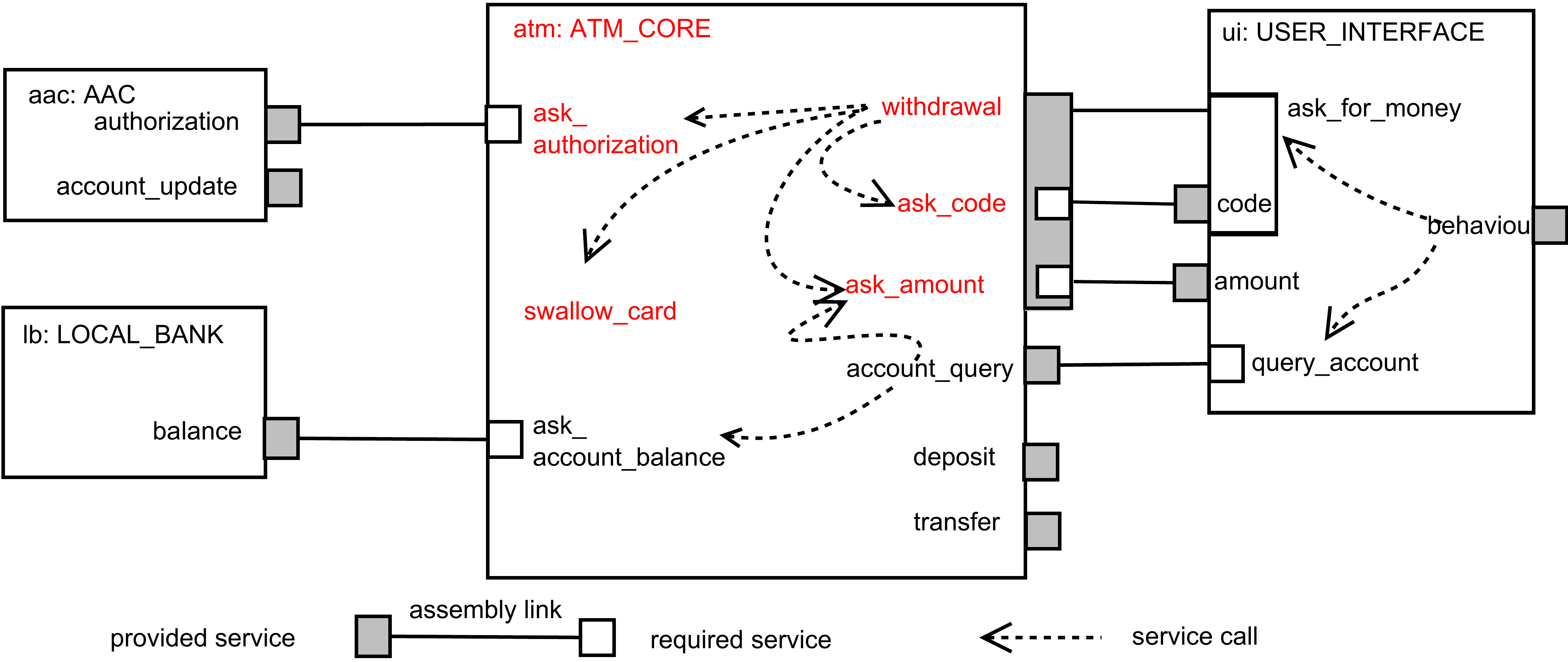}}
  \caption{A component assembly for the ATM System\label{atm.model}}
\end{figure}

Let us illustrate \kmelia by an ATM case study.
This ATM delivers standard services such as withdrawal via a user interface. 
Its \kmelia specification is built on an assembly of four components as depicted in Figure~\ref{atm.model}.
The central \K{ATM\_CORE} handles the ATM bank services;
the \K{USER\_INTERFACE} component controls the user access;
the \K{AAC} component stands for the bank management and
the \K{LOCAL\_BANK} component holds the local management access.
Components are pairwise linked: a required service is \textsl{achieved} by the provided service it is linked to.
An assembly link is a correspondence between a required service and a provided one according to mapping relations (names, context, messages, subservices).
It materialises the support for \emph{assembly contracts}.

A \kmelia component is described by an interface, a state space and service descriptions.
The interface declares the services which are provided or required by the component.
The state space is a set of variables constrained by an invariant.
In Listing~\ref{spec:kmelia:atm:core:state} the \K{ATM_CORE} state space includes an ATM name, an identifier,
a set of swallowed cards and the available notes.
The \K{CashCard} data type is defined in the user-defined library \K{ATMLIB}.
The \K{obs} prefix denotes a variable with a read-only access for a linked client service.
The invariant predicate states that there is enough cash to proceed a transaction and the "bad" card container is not full. 

\begin{multicols}{2}
\specKmeliaFirst{ATM\_CORE}{spec:kmelia:atm:core:state}{\repSPEC/ATM_CORE.cmp}{14-23}
\specKmeliaSecond{\repSPEC/ATM_CORE.cmp}{24-34}
\end{multicols}

A service may be a non-trivial entity with a state and a dynamic behaviour.
A service may also declare required and provided subservices.
All these elements are involved in the \emph{service contract}.
The service behaviour defines via an \textit{extended labelled transition system} (eLTS) the order in which the service performs its actions.
Communication actions are primitives for synchronous interactions between services.
The \K{withdraw} service achieves a withdrawal on a cash card, under some controls.
Listing~\ref{spec:kmelia:atm:core:withdraw} illustrates its declaration.

\begin{multicols}{2}
\specKmelia{of ATM services}{spec:kmelia:atm:core:withdraw}{\repSPEC/withdrawR.srv}{1-25}
\noindent
\small
The corresponding required service \K{ask_for_money} is defined in the \K{USER\_INTERFACE} component.
\normalsize
\begin{lstlisting}[language=Kmelia,basicstyle={\tiny
    \sffamily},frame=trBL,framesep=2pt]
required ask_for_money (card : CashCard) : Boolean
Interface
  subprovides : {code}
    //provided to the callee

  Virtual Variables
     dispensable : Boolean;
     // assume this observable information

  Virtual Invariant true
  Pre dispensable

  //No LTS

  Post not (Result) implies dispensable
   //dispensable may evolve in the other case

End
\end{lstlisting}
\end{multicols}

A required service may have a full service specification in \kmelia,
especially if it sets assumptions on any provider service via a \textit{virtual context}.
This allows to define service contracts separately from assembly contracts and
to improve the property verification locality.
The context mapping of the \texttt{lwith} link in Listing~\ref{atm:ass:link},
shows how the virtual context of the required service is "instantiated" by an actual context of the provider service.

\begin{lstlisting}[language=Kmelia,basicstyle={\scriptsize
    \sffamily},frame=trBL,framesep=2pt,caption="ATM Assembly links",label=atm:ass:link]
Assembly
    Components       atm:ATM_CORE;      ui:USER_INTERFACE
    Links ////////////assembly links//////////
    @lwith: p-r atm.withdrawal ui.ask_for_money
         context mapping  //a kind of explicit adaptation
           ui.dispensable = atm.available_notes >= 0
         sublinks : {lcode,lamount}
      @lamount: r-p atm.ask_amount ui.amount
      @lcode: r-p atm.ask_code ui.code
	// the service withdrawal of the ATM_CORE is connected to
	// the service ask_for_money required by the USER_INTERFACE
\end{lstlisting}

\begin{multicols}{2}
The withdrawal behaviour starts with an identification step: card insertion, password control, authentication by ACD/ATM Controller (AAC).
If the AAC accepts the transaction, the ATM asks for the amount of cash, otherwise the card is ejected and the withdrawal transaction ends.
The given amount is compared with the current card policy limit.
When the allowed amount is lower than the requested one or if the current ATM cash is not sufficient, the ATM asks again for the amount of cash.
Otherwise the ATM asks the AAC to process the transaction, updates the card limit, delivers the cash and prints a receipt when possible, and
the withdrawal transaction ends after a card ejection.
Two actions (\K{debitCard}, \K{ejectCard}) represent functions defined by the specifier in the user-defined \K{ATMLIB} library while \K{display} is a predefined function in \kmelia.

\includegraphics[width=0.48\textwidth, height=9cm]{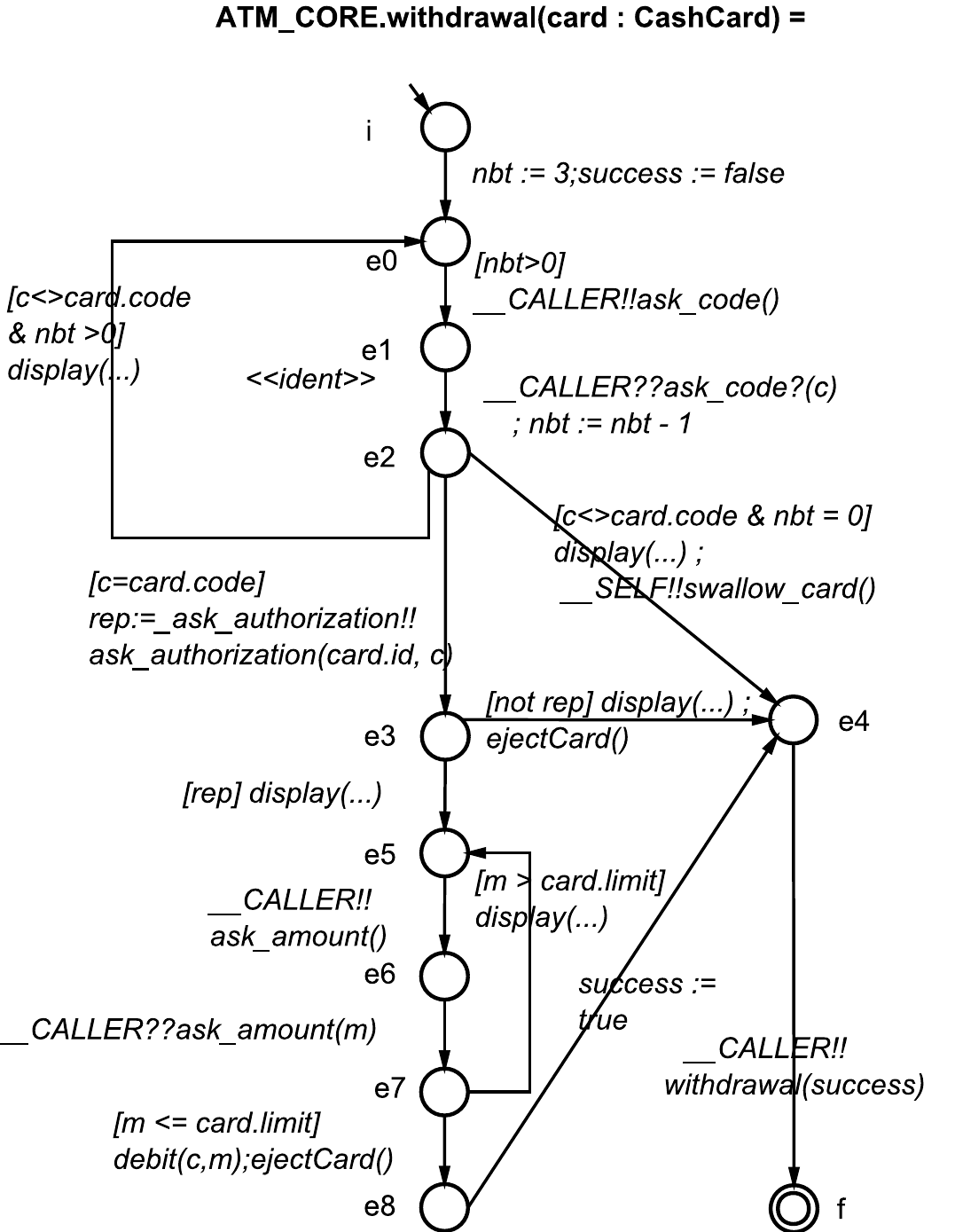}
\end{multicols}

\section{Checking Contracts in Kmelia}
\label{verification}

\begin{figure}[!ht]
\begin{center}
\includegraphics[width=0.7\linewidth]{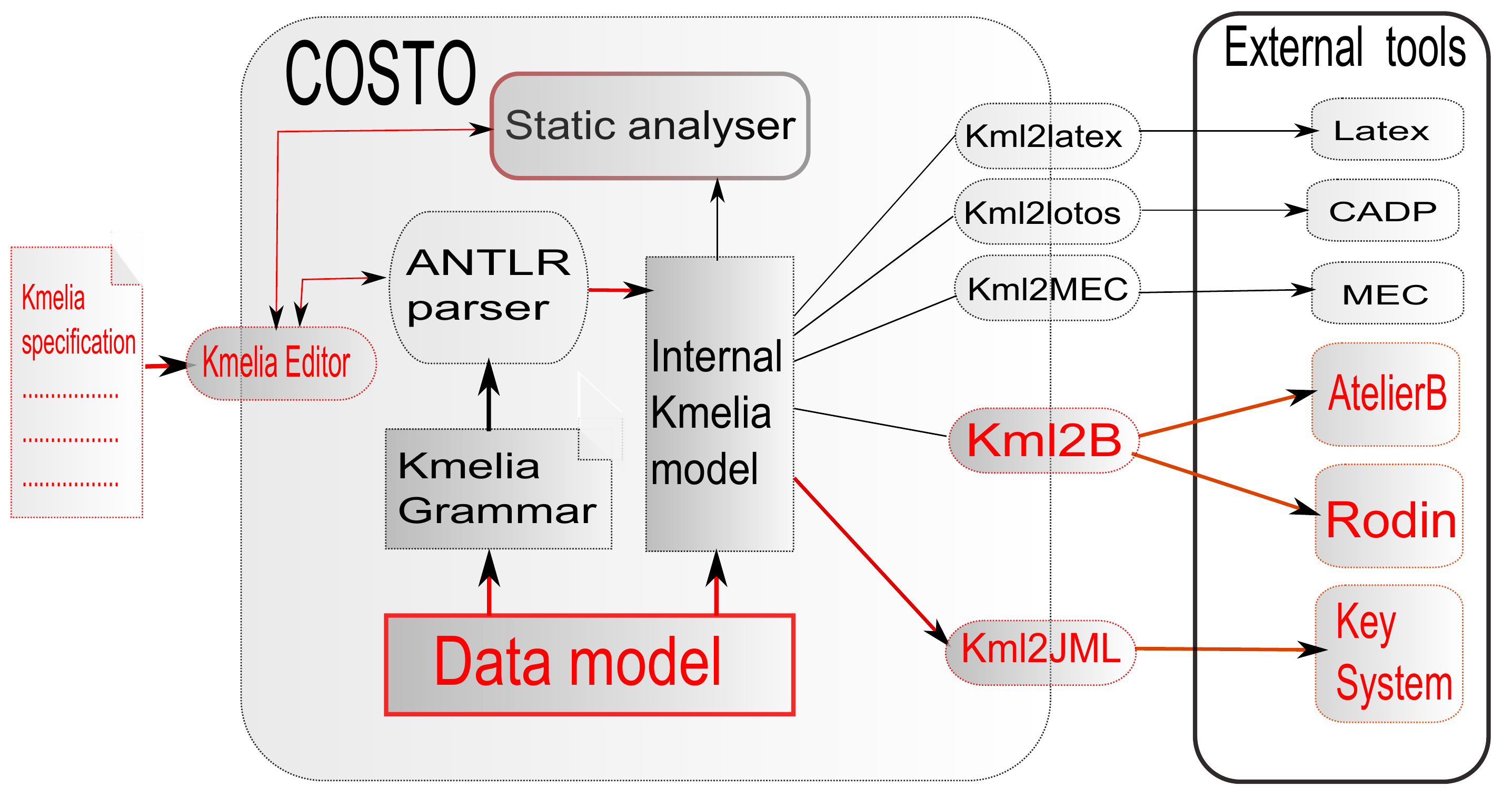}
 \caption{\textsf{COSTO} Framework Overview}\label{costokey}
 \end{center}
\end{figure}

The verification process is supported by a set of tools integrated into the
\textsf{COSTO} (COmponent Study TOolbox) platform which is a set of Eclipse-based plugins~\cite{aaaProvecs07} we developed to support the specification and analysis of \kmelia component systems. \textsf{COSTO} manages the \kmelia specifications and handles the verification of the primary properties (syntactic analysis, type checking, static analysis, ...) as depicted in Figure~\ref{costokey}. The verifications of complex properties such as deadlock freeness, component or assembly consistency are delegated to other  appropriate external tools.
Let us assume here that the static verifications (syntactic, type, well-formedness checking) are already performed by the \textsf{COSTO} tool;
we show how other verification tools are used to check contracts.

\subsection{Checking a service contract: functional correctness}
\label{chk:service:contracts}
\label{chk:service:contracts:functional}

The basic idea is to \textit{evaluate} all the paths of a service behaviour ($\mathcal B$) and to determine whether they are compliant with the post-condition or not.
Actually this is a non-trivial problem similar to the one of model-checking code. 
To prove this property we investigated B tools, including \emph{ProB} a model checker for B.
We had to turn back to more appropriate tools because B tools needed additional material to prove loop invariants and \emph{ProB} was not powerful enough.
In this section we present an investigation using the \key\footnote{\url{http://www.key-project.org}} tool~\cite{KeYBook2007}.
\key accepts JML specifications as input; therefore
we defined a process to compute JML specifications from \kmelia services.
An ongoing plugin called \textsf{Kml2Jml} implements this process (Figure~\ref{costokey}).
Each \kmelia component \K{C} is translated to a Java class \JJ{C.java} where:
each provided service of \K{C} becomes a method of the \JJ{C.java} class
and each required service of \K{C} becomes a method of a \textit{virtual} component class denoted by an instance variable \JJ{vc : VC} in the \JJ{C} class.
The LTS that specifies $\mathcal B$ is translated in two steps into a Java code. 
The translation is not straightforward, due to the gap between the LTS structure and the structured programming control structures of Java.
The first step introduces the control structure by converting the LTS into a syntax tree, which is an extension of the well-known regular expressions formalism.
The second step computes the Java code from the syntax tree.

\noindent
\textbf{Algorithm (step1): From LTS to Syntax tree}
Let $L(\mathcal B)$ bet the set of possible behaviours of a service $srv$.
Since $\mathcal B$ can be seen as an automaton,
the syntax tree $E$ such that $L(E) = L(\mathcal B)$ is generated by the algorithm of \emph{McNaughton/Yamada} (cf. \emph{Kleene's theorem} in \cite{mny}).

\noindent
\textbf{Algorithm (step2): From syntax tree to Java} It is straightforward from the previously obtained syntax tree. The main idea is to transform the product ($.$) as a sequence operator (; in Java), the Union $(+)$ as a conditional structure (if else ...)\footnote{Note that even non-deterministic choice in LTS can be modeled by an if-else construction over an abstract variable that could be refined later by developer}, and the Kleene star $(*)$ as a recursive method modeling $E_*$. The body of this method describes a statement block repeated in the LTS. The resulting Java code annotated with JML specifications is checked using the \key tool.

\paragraph{Example.}
Let us check the functional correctness of the \textit{withdrawal} provided service of the \BN{ATM_Core} component.
Applied to the Java representation of the \textit{withdrawal} service, the \key tool analysis revealed some errors. 
As an example, the post-condition of \textit{withdrawal} was not satisfied:
the error was due to the addition of \JJ{amount} to \JJ{availaible_notes} instead of subtracting it.
After correcting this mistake, and regenerating the Java code, \key  proved it correct automatically with $654$ \textit{symbolic states} and $18$ \textit{path conditions}.

\subsection{Checking a component contract: component consistency}
\label{chk:service:contracts:consistency}

Here the deal is to reuse B tools like Atelier-B\footnote{\url{http://www.atelierb.eu/}} and Rodin\footnote{\url{http://rodin-b-sharp.sourceforge.net}} because
the B provers are appropriate to prove that kind of property, 
considering the fact that most of the \kmelia data types and expression are translatable in B.
We developed a plugin named \texttt{Kml2B} in \textsf{COSTO}
 (Figure~\ref{costokey}) that extracts (Event-)B specifications.
For each \kmelia component $C$, an (Event-)B model called \BN{C} is built.
Its state space is extracted  from the component's one. The  provided services $srv_i$ in $C$ are translated into \BN{srv_i} operations within the \BN{C} model. The extracted specification is imported and checked in  Atelier-B or Rodin.
The B tools enables the verification of invariant consistency at the \kmelia level.
The full translation procedure is explained in~\cite{cal10}.

\paragraph{Example.}
After extracting (Event-)B models by running the \textsf{Kml2B} plugin, the \BN{ATM_Core} model is used to prove the preservation of invariant by its provided services. We proved consequently the consistency of the invariant component. However, if the post-condition is modified as $available\_notes <= old(available\_notes)$ then the invariant $ available\_notes >= 0 $ is not  preserved anymore. This error was easily detected with the \B tools.

\subsection{Checking assembly contracts}
\label{chk:assembly:contracts}

Checking an assembly contract engages four verifications steps:
(i) the matching of the service signatures (up to parameter renaming),
(ii) the service dependency consistency,
(iii) the matching of the service pre/post-conditions and
(iv) the behavioural compatibility of services.
Step (i) and (ii) are performed by checking static interoperability.

\subsubsection{Checking the static interoperability}
\label{chk:assembly:contracts:static}
This verification includes: type checking, signature matching, component and service interfaces structure matching, observability rules, and service availability (requirements, subservices). The COSTO tool performs these analysis  by using simple correspondence checking algorithms, graph algorithms and standard typing algorithms.
Static analysis of these contracts helps to detect some incompatibilities; therefore the component designer may correct its component at design time. This corresponds to the "Static Analysis" step in Figure~\ref{process}. The reader can find  more details of this analysis in~\cite{facs09}.

\begin{figure}[!ht]
\includegraphics[width=1\linewidth]{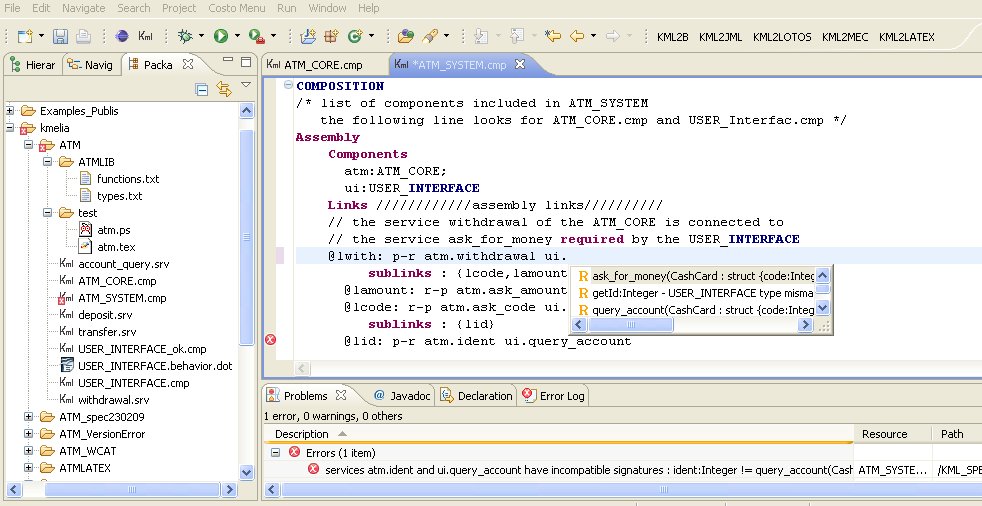}
 \caption{Error detection and smart completion in COSTO/\kmelia\label{figure:costo_errors}\label{figure:costo_completion}}
\end{figure}

\paragraph{Example.}
Figure~\ref{figure:costo_errors} shows the \kmelia editor in the Eclipse IDE,  and a sample of the kind of errors (typing, observability, incompleteness of the mapping) that are detected.
Besides standard completion, the editor supports smart completion in the case of assembly links. In Figure \ref{figure:costo_completion},  only required services defined in  the \K{User_Interface}  component type are proposed and the user is warned that some of them do not match the exact signature of the provided service \K{withdraw} which is defined in the \K{ATM_Core} component type.

\subsubsection{Checking the service contracts compliance}
\label{chk:assembly:contracts:compliancy}

Based on an assembly link, the main issue is to decide whether the provided service matches with the required service it is linked to.
The matching condition is: \textit{the pre-condition of required service $Req$ is stronger than the one of provided service $Prov$ and the post-condition of $Req$  is weaker than the  one of $Prov$}.
In term of B proof obligations this property is rewritten as: the provided service refines the required service (considering the adaptation defined
in the context mapping of the link).
The refinement relation leads to the generation of specific proof obligations in (Event-)B.
In practice we reuse the work presented in Section~\ref{chk:service:contracts:consistency}.
We extend it by generating B machines for the required services and for the refinement relation.
For each required service $Req$ in a component $C$, one (Event-)B model \BN{Req} is created before checking the consistency of the virtual context of the service $Req$. The same (Event-)B model is refined by \BN{Req\_Prov\_Ref}.
The state space of the machine \BN {Req} is obtained by translating the virtual context of service $Req$, and the operation \BN{req} is the translation of the service $Req$.
The full details of the translation schema and the proof obligations are available in~\cite{fesca10} while the \textsf{Kml2B} plugin is introduced
in~\cite{cal10}.

\paragraph{Example.}
The analysis of the assembly link \K{lwith} between the required service \K{ask_for_money} and the provided service \K{withdraw_ref} with the \BN{AtelierB} reveals some errors that were introduced intentionally in the specification of \K{ask_for_money} for experimental purpose.
The post-condition (\K{not(result) implies} \K{not (dispensable)}) means that if $result$ is $false$ then the $available\_notes$ is less than $0$ which can not be deduced from the \K{withdraw_ref} post-condition. Then the  service contract compliance $ (Post(waithdraw\_ref) \Rightarrow$ $ Post$ $(ask\_for\_money)) $ is not fulfilled.  After correcting the error, the resulting B machines  generated $28$ proof obligations which were all proved by the \BN{AtelierB} prover in  \emph{Automatic} mode.

\subsubsection{Checking behavioural compatibility}
\label{chk:assembly:contracts:behavioural}

This verification focus on the synchronous communication actions between services (start/end of services, send/receive message) defined in the \kmelia model.
{Checking behavioural compatibility is a widely studied topic \cite{YellinStrom97,Attie:2003:CMC,BraccialiBC05}.
It often relies on checking the behaviour of a (component-based) system through the construction of a finite state automaton.
We adopt a pairwise verification approach that avoids state explosion as described in \cite{Attie:2003:CMC}. 
Hence for each assembly link, including the sublinks that share the same communication channel, the behavioural compatibility verification is applied.
Instead of developping the checker we turned to existing model checkers
because ensuring dynamic behavioural compatibility is usual target property of communicating processes and transition systems. 
Currently we target MEC and CADP tools.
In order to exploit the CADP tools~\cite{FeGaKe+96}, we encode the \kmelia components into \Lotos processes which are the input of the CADP tools. The behavioural compatibility is based on communication between processes.
A plugin named \textsf{Kml2Lotos} have been developed in a previous work \cite{sc06}. The resulting \Lotos process can be checked using CADP tool. An alternative solution based on MEC model checker have been also experimented.

\paragraph{Example.}
The  experimentations led to detect message inconsistencies.
The error made by the specifier was to put a message reception in a loop for one service and a single sent message in the communication service. The deadlock was reached in case of a second pass in the loop. The MEC translator \textsf{Kml2Mec} and a full experimentation with MEC can be found in~\cite{mosim06}.

\section{Related Works}
\label{related}
Using contracts for components is not a new topic.
However to the best of our knowledge the related approaches do not integrate contracts the way we propose.
Contracts for component have been described in~\cite{Schmidt1}.  That
proposal considers functional and extra-functional contracts and
dynamic behaviours to provide trust-by-contract components. However
the main issue of that work is software quality; the proof of the
contracts is not treated at the design level.
Beugnard et al.\cite{Beugnard} investigated a typology of component contracts and  classified contracts in four levels. Basically \textit{syntactic contracts (i)} are taken into account by all component models. The more relevant semantic constraints such as \textit{behavioural contracts (ii)} and \textit{synchronisation contracts (iii)} are encountered in specific component models.
Finally the \textit{quality of service (iv)} is often delegated to runtime models. 

In ConFract~\cite{confract} the contracts are independent
entities associated to several participants, while \kmelia attaches them mainly to  services and links. The ConFract contracts support a rely/guarantee
mechanism with respect to the vertical composition of Fractal
components~\cite{FractalSPE06}.
The executable assertions language CCL-J enables to express
specifications at the interface level and the component levels. 
In the case of CCL-J, when a method of an interface is called, the contract controller is notified and it applies the checking rules. As for the pre-conditions, the post-conditions and the method invariants of all contracts "are checked at
runtime". CCL-J is used to validate the contracting mechanisms of ConFract but CCL-J is much simpler than JML in terms of available constructs.
In~\cite{DBLP:conf/cbsq/ReussnerPS03} the definition of
Meyer's contracts and subcontracts is assumed, which led to rules
similar to those of \kmelia. But the interpretation of pre-conditions
and post-conditions is done in terms of call sequences rather than in
logical predicates.  This relies on behavioural contracts rather than
functional contracts.  In \kmelia, the behavioural contracts are treated
separately using behaviour compatibility rules~\cite{sc06}.
The SOFA component model and its behaviour protocol formalism \cite{sofa}, based on regular expressions, allow the designer to verify the conformance of a component's implementation to its specification; this verification is done at runtime.
But no service contracts compliance is handled.

Architecture Description Languages represent software architectures in terms of components and their overall interconnection structure. Many of these languages support formal notations to specify components and connectors behaviours. For
example, Wright \cite{wright} and Darwin \cite{basa} use CSP-based notations. These formalisms allow to verify correctness of component assemblies, to check properties such as deadlock freedom. However most of the works applying formal verifications in ADLs focus on component interactions, but very few studies addressed the contract issue using pre/post-conditions.
Apart from the  \textit{syntactic contracts} level (i), the \textit{behavioural contracts (ii)}  and the \textit{synchronisation contracts (iii)} are also proved at design time in \kmelia. We do not deal with further constraints such as quality of service, because they depend on data known only at runtime.

Contracts and services have been studied in the context of service composition.
From a service composition point of view \textit{e.g.} BPEL, the behavioural aspect is dominant \cite{Beek2007}.
Considering only the formal models, composition is mainly based on automata, Petri nets and process algebra, as illustrated by the orchestration calculus of Mazzara and Lanese \cite{MazzaraL06};
therefore the contracts focus mainly on dynamic compatibility.
Conversely the contracts (in the sense of \textit{design-by-contract}) are taken into account in \cite{Milanovic05} (using abstract machines) but not the dynamic behaviour.
\kmelia cares of both aspects.
In \cite{brogiJLAP2010}, the contract is supported at four levels  (signature, quality of service, ontology, behaviour) but none of them handle the functional contract. 
The component architecture (SCA) approaches \cite{DingCL08} emphasize the
service concept, like \kmelia does; but  contract features are not introduced yet in SCA.

\section{Conclusion}
\label{conclusion}

We have presented how a set of correctness properties of components may be guaranteed by stating and verifying contracts at the level of services, components and assemblies. We have illustrated the idea through the \kmelia model which is equiped with a rich data language that enables to incorporate pre/post-conditions at a service level, invariant at a component level, and behavioural contract at an assembly level. Consequently, property verification is achieved by checking the contracts at the different levels. The automation of the process is undertaken by considering extractions from the \kmelia specification language to generate the specifications in the input language of existing tools such as theorem-provers or model-checkers depending on the targeted properties.
The use of a multilevel contracts makes it easy to define interoperability policy. For instance static interoperability exploits low level pre/post-conditions and helps us to check the correctness of assemblies. This may be generalised to assemblies of heterogeneous components, provided that a standard pre/post-condition mechanism is defined and respected.

CBSE lacks standard practices in order to raise a large-scale, open use of components. The road to a wide spread component-based software engineering is simplicity, ease of use, availability of well-defined, standard, free and useful components and interfaces. The Unix operating system is a convincing example that makes the proof of the concept, at a different level; the simple use of Unix \textsf{.h} header interfaces, the simple combination of Unix commands and options, the simple use of unstructured files, the conformance of the standard interfaces including  network levels, are recognised as the main points for the development of operating system components that make the success of the Unix software family.      
We expect that first order logic integrated in high-level programming languages or operating systems as the use of script languages, can play a similar role of interface standardisation for CBSE. We are working in this direction via the reuse and the extension of existing standard relational database languages (the \textsf{SQL} family) which are already integrated in various operating system features.\\
\textit{Perspectives.}
A short term perspective of our work is to make the tools used at different levels more integrated with helpful feedback into the \kmelia specifications.
We are working on a translation of a subset of \kmelia into
 the Fractal component model which has a Java execution environment but
 lacks property verification means.
We expect to favour interoperability between the models and also to find some simulation facilities that will be complementary with the formal analysis aspect provided by \kmelia.

\medskip
\noindent
\textbf{Acknowledgements} We thank the anonymous reviewers for their comments which helped us to improve the paper, and our colleague Gilles Ardourel for helping in the preparation of this final version. 

\small
\noindent 

\bibliographystyle{eptcs} 
\bibliography{wcsi10}
\end{document}